\documentstyle[preprint,aps]{revtex}

\input epsf
\tightenlines

\begin{document}

\draft
\title{Deutsch-Jozsa algorithm as a test of quantum computation}
\author{ David~Collins, K.~W.~Kim and W.~C.~Holton}
\address{ Department of Electrical and Computer Engineering,
          Box 7911,
          232 Daniels Hall,
	  North Carolina State University, Raleigh, 
	  North Carolina 27695-7911}
\date{\today}
	
\maketitle	

\begin{abstract}
 A redundancy in the existing Deutsch-Jozsa quantum algorithm is removed and
 a refined algorithm, which reduces the size of the register and simplifies 
the function evaluation, is proposed. 
The refined version allows a simpler analysis of the use of entanglement 
between the qubits in the algorithm and provides criteria for deciding when 
the Deutsch-Jozsa algorithm constitutes a meaningful test of quantum 
computation. 
\end{abstract}

\pacs{03.67.Lx, 03.65.-w}


Quantum computation has emerged in the past decade as a potentially very 
powerful way to  solve certain problems. The idea is to store information 
in states of a quantum system, manipulate these via unitary transformations
 and extract useful information from the resulting state. The use of key 
features of quantum mechanics such as superposition of states and quantum 
entanglement enables exponential speedups in the solution of certain 
problems. Within this framework computational schemes, algorithms, and 
error correction have been developed \cite{ekertjozsa,bennett}. However, 
practical implementation, which requires precise control of quantum 
systems, remains beset by difficulties \cite{preskill}. Despite skepticism 
about the feasibility of quantum computation several groups claim to have 
demonstrated experimentally the operation of quantum gates \cite{monroe}, 
error correction \cite{cory} and simple algorithms 
\cite{chuang,jones,chuangsearch}. 

To date most experimental work on quantum algorithms has been directed 
toward implementations of the Deutsch-Jozsa algorithm, which provides a 
fertile ground for illustrating the key features of quantum computation 
\cite{deutsch,chuangyamamoto,vedral,cleve,jozsa}. Our purpose is to 
investigate the Deutsch-Jozsa algorithm to determine when it is a 
meaningful test of quantum computation. We shall focus on the algorithm's 
use of entanglement, which gives quantum computation its power \cite{jozsa}.

The Deutsch problem \cite{deutsch,cleve} considers certain global 
properties of functions on $N$-bit binary numbers. Denote the set of all 
such numbers by ${\bf X}_N := \left\{ x_N x_{N-1} \ldots x_1 x_0 \; \; | \;
 \; x_m = 0,1 \right\}$. A function $f: {\bf X}_N \rightarrow \left\{0,1 
\right\}$ is called {\it balanced} if the number of times it returns $0$ is
 equal to the number of times it returns $1$ as the argument ranges over 
${\bf X}_N$. The Deutsch problem is to take a given function which is known
 to be either balanced or constant and to determine which type it is. A
classical algorithm which would answer this with certainty would require 
that $f$ be evaluated for $2^{N-1} + 1$ values of its argument and thus 
grows exponentially with input size.
The Deutsch problem may be solved by a quantum algorithm 
\cite{chuang,jones,deutsch,cleve} which only requires one ``evaluation'' 
of $f$. This requires: (1) A well defined physical system, called the 
total register, to be used for storing and retrieving information. (2) A 
sequence of unitary transformations to be enacted on the total register in 
such a way as to produce an answer to the problem.

The total register in the existing algorithm consists of an $N$ qubit 
{\em control register}, which is generally used for storing the function 
arguments, plus a one qubit {\em function register}, which is used for 
function evaluation. The Hilbert space for the control register will be 
denoted ${\cal H}_c$ and that for the function register ${\cal H}_f$. Thus
 the total register in the existing algorithm is 
\begin{equation}
 {\cal H}_{\text{tot}} = {\cal H}_c \otimes {\cal H}_f.
 \label{eq:totreg}
\end{equation}
Basis elements for the control register will be denoted
\begin{equation}
 \left| x \right>_c \equiv 
 \left| x_{N-1} \ldots x_0 \right>_c \equiv
 \left| x_{N-1}\right>_c 
 \ldots 
 \left| x_0\right>_c
 \label{eq:controlreg}
\end{equation}
where $\left\{ \left| x_m \right>_c \; \; | \; \; x_m = 0,1 \right\}$ are 
orthonormal basis sets for distinguishable two-state systems. Similarly 
basis elements for the function register are denoted $\left| y \right>_f$ 
where $y = 0,1$. Note that equation (\ref{eq:controlreg}) provides a 
one-to-one correspondence between ${\bf X}_N$ (function arguments) and 
basis elements of the control register.
Thus any physical implementation of the existing algorithm, such as those 
accomplished using an NMR quantum computer \cite{chuang,jones}, requires a 
two-state system for the function register in addition to those for the 
control register.    

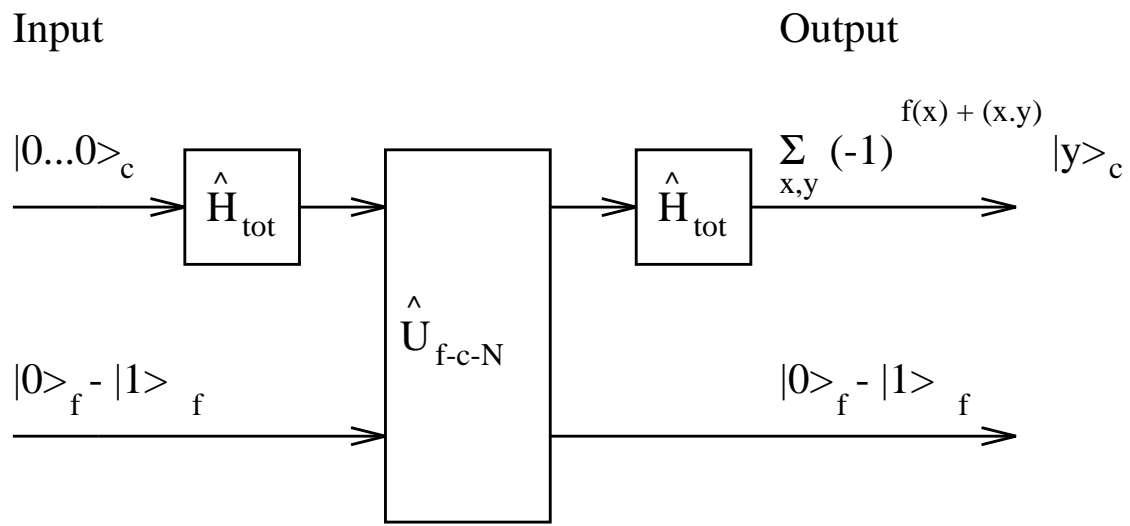
\begin{figure}[h]
                     \caption{The existing Deutsch-Jozsa algorithm. 
$\hat{H}_{\text{tot}}$ represents a Hadamard transformation applied to each
 qubit of the control register. $\hat{U}_{f-c-N}$ represents the operation
 of the $f$-controlled-NOT gate. Note that here 
$x.y := x_{N-1}y_{N-1} + \ldots x_0 y_0$. Summation is over all elements of
 ${\bf X}_N $.} 
		     \label{pic:existingdj}
\end{figure}

The existing state-of-the-art quantum algorithm (see the articles by 
Cleve {\it et. al.} \cite{cleve} and Jones and Mosca \cite{jones}) for the 
Deutsch problem \cite{chuang,jones,cleve} makes use of the total register 
given in equation (\ref{eq:totreg}), and follows the scheme illustrated in 
figure \ref{pic:existingdj}.   
The two gates used are: 
\begin{itemize}
 \item[1.] A Hadamard transformation applied to each qubit of the control 
register, $\hat{H}_{\text{tot}}~:=~\hat{H}~\otimes~\ldots~\otimes~\hat{H}$,
 where 

\begin{equation}
 \hat{H} := {\frac{1}{\sqrt{2}}
           \left( \begin{array}{cr}
                         1 & 1 \\
                         1 & -1
           \end{array} \right) }.
\end{equation}
This transforms the input state $\left| 0 \ldots 0 \right>_c$ into an equal 
superposition over all basis elements given in 
equation~(\ref{eq:controlreg}), thereby preparing the way for a 
simultaneous evaluation of $f$ over all possible arguments.

 \item[2.] The $f$-controlled-NOT gate, $\hat{U}_{f-c-N}$, which performs 
the function evaluation, whose operation on an orthonormal basis of 
${\cal H}_{\text{tot}}$ is defined as

\begin{equation}
 \hat{U}_{f-c-N}\left| x \right>_c \left| y \right>_f := 
                         \left| x \right>_c 
                         \left| y \oplus f(x) \right>_f 
\end{equation}
and is extended linearly to all elements of ${\cal H}_{\text{tot}}$.
\end{itemize}
The final step is evaluation of the expectation value of 
$\left| 0 \ldots 0 \right>_c \left< 0 \ldots 0 \right|_c$ on the control 
register. If the result is $0$ then $f$  is 
balanced and if it is $1$ then $f$  is constant. It is easily shown 
that this algorithm exhibits polynomial growth in input size.

Clearly the only point at which entanglement between any of the qubits of
 the total register could occur is during the function evaluation. However,
 it is also apparent that the function register is in the state 
$\frac{1}{\sqrt{2}}\left[ \left| 0 \right>_f - \left| 1 \right>_f \right]$ 
before and after this step. In fact it is easily shown that 

\begin{equation}
 \hat{U}_{f-c-N}  \left| x \right>_c
 \frac{1}{\sqrt{2}}\left[ \left| 0 \right>_f - \left| 1 \right>_f \right] =
  \left( -1 \right)^{f(x)}  \left| x \right>_c
 \frac{1}{\sqrt{2}}\left[ \left| 0 \right>_f - \left| 1 \right>_f \right]
 \label{eq:noentang}
\end{equation}
which implies that {\em if the function register of the input state is 
restricted to the subspace spanned by $\left[ \left| 0 \right>_f - \left| 
1 \right>_f \right]$ then there is no entanglement between the control and 
the function registers in the output of the $\hat{U}_{f-c-N}$}. Thus there 
is no need for any coupling between the two-state systems comprising the 
control register and that comprising the function register. In the existing
 algorithm the function register is completely redundant.

This suggests that the algorithm should be modified by eliminating the 
function register while retaining  the control register of the previous 
algorithm. Thus the total register is ${\cal H}_c$. 

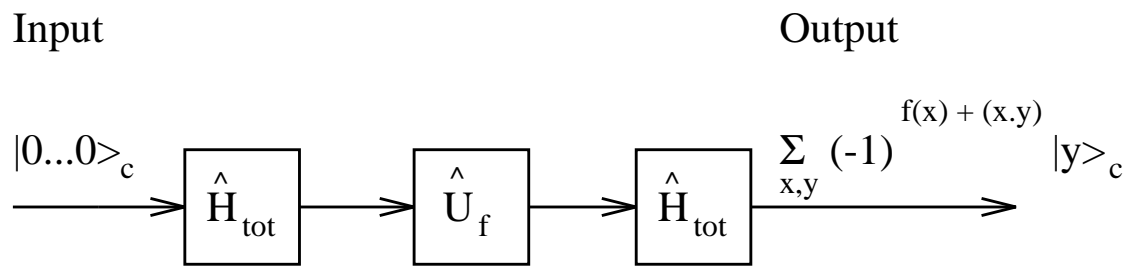
\begin{figure}[h] 
                     \caption{The refined Deutsch-Jozsa algorithm. Notation
 is the same as in figure \ref{pic:existingdj} with the $f$-controlled gate
 $\hat{U}_f$ replacing the $f$-controlled-NOT gate.} 
		     \label{pic:djref}
\end{figure}

Equation (\ref{eq:noentang}) suggests that the function evaluation can be 
carried out via the {\em $f$-controlled gate} whose operation on the basis 
elements of the control register is defined as

\begin{equation}
  \hat{U}_f \left| x \right>_c := \left( -1\right)^{f(x)} \left| x \right>_c
  \label{eq:fcontrdef}
\end{equation}
and which is extended linearly to all elements of ${\cal H}_c$.
Indeed it is easily seen that in the existing version of the algorithm the 
effect of $\hat{U}_{f-c-N}$ is identical to that of $\hat{U}_f \otimes 
\hat{I}_f$  where $\hat{I}_f$ is the identity on ${\cal H}_f$. Note that 
this is invalid for the most general element of ${\cal H}_{\text{tot}}$ in 
the existing algorithm but is true whenever the input state of the function
 register is in the subspace spanned by 
$\left[ \left| 0 \right>_f - \left| 1 \right>_f \right]$. Clearly 
$\hat{U}_f$ satisfies the requirement that a gate must be a unitary 
operator.
The refined algorithm follows a similar pattern of operations as the 
existing algorithm. A schematic form is provided in figure \ref{pic:djref}.
The refined algorithm requires one qubit fewer than the existing algorithm.
 Consequently physical implementation requires one fewer two-state system. 
Thus the experimental demonstrations using NMR systems \cite{chuang,jones} 
could have been carried out using a single  spin one-half system instead of
 the two coupled spin one-half systems that were actually used.

Typically quantum computers achieve their efficiency by utilizing
 entanglement between the various qubits of the total register \cite{jozsa}.
The above discussion shows that for both the existing and refined 
Deutsch-Jozsa algorithms the only possibility for entanglement is amongst 
the qubits of the control register through the action of the $f$-controlled
 gate. Furthermore, whether or not entanglement occurs depends on the form 
of $f$. For example, equation (\ref{eq:fcontrdef}) shows that if $f$ is 
constant then $\hat{U}_f = \pm \hat{I}_c$ and it does not cause any 
entanglement between the qubits of the control register. The need for 
entanglement for any balanced function requires a straightforward but 
tedious analysis of $\hat{U}_f$; for some balanced functions it causes 
entanglement and for others not.
We shall demand that for a given value of $N$ the quantum computer 
(i.\ e.\ physical system) must be capable of solving the Deutsch problem 
for all admissible functions. We thus investigate whether a given qubit of 
the control register remains unentangled (after operation of $\hat{U}_f$) 
from the remaining qubits for all possible balanced functions. If so it can
 remain uncoupled from the rest of the register and can be implemented on a
 completely isolated quantum system. 

For $N=1$ entanglement is clearly not an issue. For $N=2$ it can be shown 
that (from here onwards the arguments of the $f$ are expressed in decimal 
form) 

\begin{equation}
 \hat{U}_f = \hat{U}_1 \otimes \hat{U}_0
 \label{eq:factor2} 
\end{equation} 
where $\hat{U}_m$ operates on the $\left|{x_m} \right>_c$ qubit and, with 
respect to the basis 
$\left\{\left| 0 \right>_c, \left| 1 \right>_c \right\}$,

\begin{equation}
 \hat{U}_1 = \left( \begin{array}{cc}
                         1 & 0 \\
                         0 & \left( -1 \right)^{f(0) + f(2)}
                         \end{array} \right)
 \label{eq:factor2a} 
\end{equation}
and 

\begin{equation}
 \hat{U}_0 =  \left( -1 \right)^{f(0)} 
               \left( \begin{array}{cc}
                         1 & 0 \\
                         0 & \left( -1 \right)^{f(0) + f(1)}
                         \end{array} \right) 
  \label{eq:factor2b} 
\end{equation}
where we have used the fact that for any balanced function

\begin{equation}
 \left( -1 \right)^{f(3)} = \left( -1 \right)^{f(0) + f(1) + f(2)}. 
 \label{eq:frelation}
\end{equation}
Note that $\hat{U}_1$ and $\hat{U}_0$ are unitary.
Thus for $N=2$ the $f$-controlled gate does not cause entanglement between 
the qubits of the control register. 

Now consider $N>2$ and assume that the $\left| x_0 \right>_c$ qubit is not 
entangled by the $f$-controlled gate. Thus 

\begin{equation}
 \hat{U}_f = \hat{U}' \otimes \hat{U}_0
\end{equation}
where $\hat{U}'$ operates on the qubits 
$\left| x_{N-1} \ldots x_1 \right>_c$ and 
$\hat{U}_0$ operates on $\left| x_0 \right>_c$.
It is easily shown that, with respect to the basis 
$\left\{\left| 0 \right>_c, \left| 1 \right>_c \right\}$, 

\begin{equation}
    \hat{U}' = \left( \begin{array}{ccccc}
               a_1 & 0 & \ldots & 0 & 0 \\
               0   & a_2 & \ldots & 0 & 0 \\
               \vdots &  & \ddots &  & \vdots \\
               0 & 0 & \ldots & a_{(2^{N-1}-1)} & 0 \\
               0 & 0 & \ldots & 0 & a_{(2^{N-1})} 
              \end{array} \right) 
\end{equation}
where $a_m \neq 0$ and 

\begin{equation}
          \hat{U}_0 = \left(  \begin{array}{cc}
                        b_1 & 0 \\
                        0   & b_2 
                       \end{array} \right)
\end{equation}
where $b_m \neq 0$. This requires relationships of the form (expressing the
 arguments of the function in decimal form):

\begin{eqnarray}
 \left( -1 \right)^{f(0)} & = & a_1 b_1 \nonumber \\
 \left( -1 \right)^{f(1)} & = & a_1 b_2 \nonumber \\
 \left( -1 \right)^{f(2)} & = & a_2 b_1 \nonumber \\
 \left( -1 \right)^{f(3)} & = & a_2 b_2  
\end{eqnarray}
etc., which implies $\left( -1 \right)^{f(3)} = 
\left( -1 \right)^{f(0) + f(1) + f(2)}$ etc.. For $N > 2$ there clearly 
exist balanced functions for which this is invalid and for these the 
$f$-controlled gate entangles the $\left| x_0 \right>_c$ qubit with the 
rest of the control register. In this fashion it can be shown that for $N >
 2$ none of the control register qubits are always unentangled with the 
others (for all possible balanced functions).

Physically this implies that for $N=1 \; \mbox{or} \;  2$ the Deutsch-Jozsa
 algorithm may be carried out by using {\em uncoupled} two-state systems. 
 However, for $N > 2$ there is always some balanced function which requires 
that a given qubit of the control register be coupled to the remaining 
qubits. Thus meaningful tests of the Deutsch-Jozsa algorithm occur if and 
only if $N>2$. Note that the previous NMR demonstrations were conducted for 
the case where $N=1$ \cite{chuang,jones}.

This leaves the question of the utility of quantum computation for the 
cases in which $N \leq 2$, where the quantum algorithm still appears to 
answer the problem with just one function ``evaluation'' as opposed to 
$2^{N - 1}+1$ as required by the classical algorithm. It has usually been 
assumed that the best classical solution is to inspect and compare the 
first $2^{N -1} +1$ elements in the following representation of $f$: 
$\left( f(0), f(1), \ldots, f(2^N - 1) \right)$. However, the problem may 
be solved by checking the parity (even or odd) of the first $2^{N -1}$ 
elements in the following alternative representation: $\left( f(0)+ (f1), 
f(0)+f(2), \ldots, f(0)+f(2^N - 1), f(0)-f(1) \right)$. If they are all 
even then $f$ is constant and if any is odd $f$ is balanced. Thus for $N=1$
 the problem is answered by checking the parity of $f(0)+ f(1)$ which 
requires only one ``evaluation''. For $N=2$ the problem is answered by 
checking the parities of $f(0)+ f(1)$ and $f(0)+ f(2)$. Inspection of 
equations~(\ref{eq:factor2})-(\ref{eq:factor2b}) shows that the quantum 
algorithm decides these using the $\left| x_0\right>$ qubit 
(for $f(0)+ f(1)$) and the $\left| x_1\right>$ qubit (for $f(0)+ f(2)$) 
independently. Thus the ``one function evaluation'' is essentially two 
simultaneous evaluations on independent one-qubit computers, each of which 
carries out part of the classical algorithm. The apparent gain made by the 
quantum computer has occured only because the number of computers has been 
doubled; the method of solution is essentially classical and the problem 
can be solved just as easily with two classical computers. Thus for 
$N \leq 2$ the quantum algorithm solves the Deutsch problem in a classical 
way.

To conclude we have shown that it is possible to simplify the existing
 quantum algorithm for the Deutsch problem by eliminating the function 
register and redefining the function evaluation in terms of the 
$f$-controlled gate. We also showed that in the existing algorithm there is
 no entanglement between the control and function registers. In the refined
 quantum algorithm which we have presented, entanglement occurs between the
 qubits of the register only when $N>2$. Thus in order to utilize the full
 power of quantum computation as applied to the Deutsch problem, an 
implementation for $N>2$ is necessary.

This work was supported, in part, by the Defense Advanced Research Project 
Agency and the Office of Naval Research. We would also like to thank Gary 
Sanders for his help in preparing this article.

\pagebreak

\pagestyle{empty}

\vspace*{1in}

\begin{figure}[h] 
                     \centerline{\epsffile{fig1.eps}}
\end{figure} 

\vspace{4in}

\centering \large{Figure 1}

\pagebreak

\vspace*{1in}
\begin{figure}[h] \centerline{\epsffile{fig2.eps}}
\end{figure}

\vspace{5in}

\centering \large{Figure 2}


\begin{references}

 \bibitem{ekertjozsa} A. Ekert and R. Jozsa, Rev. Mod. Phys. {\bf 68}, 733 
(1996).
 
 \bibitem{bennett} C. Bennett,  Phys. Today, Oct 1995, 24 (1995).
 
 
 \bibitem{preskill} J. Preskill, Proc. R. Soc. London, Ser. A {\bf 454}, 
469 (1998).
 
 \bibitem{monroe} C. Monroe, D.M. Meekhof, B. E. King, W. M. Itano and 
D. J. Wineland,  Phys. Rev. Lett. {\bf 75}, 4714 (1995)
 
 \bibitem{cory} D. G. Cory, W. Mass, M. Price, E. Knill, R. Laflamme, 
W. H. Zurek, T. F. Havel and S. S. Somaroo, Report No. quant-ph/9802018.
 
   \bibitem{chuang} I. L. Chuang, L. M. K. Vandersypen, X. Zhou, 
D. W. Leung and S. Lloyd, Nature {\bf 393}, 143 (1998).
  
 \bibitem{jones}  J. A. Jones and M. Mosca,  Report No. quant-ph/9801027.

 \bibitem{chuangsearch} I. L. Chuang, N. Gershenfeld and M. Kubinec, Phys. 
Rev. Lett. {\bf 80}, 3408 (1998)  

 \bibitem{deutsch} D. Deutsch and R. Jozsa,  Proc. R. Soc. London, Ser. A, 
{\bf 439}, 553 (1992).
 
 \bibitem{chuangyamamoto} I. L. Chuang and Y. Yamamoto, Phys. Rev. A 
{\bf 52}, 3489 (1995)
 
 \bibitem{vedral} V. Vedral and M B. Plenio, Report No. quant-ph/9802065.
 
 
 \bibitem{cleve} R. Cleve, A. Ekert, C. Macciavello and M.Mosca, Proc. R. 
Soc. London, Ser. A {\bf 454}, 339 (1998).

 \bibitem{jozsa} R.Jozsa, Report No. quant-ph/9707034.
  
 

 
 \end{references}
\end{document}